\documentclass[twocolumn,showkeys,aps,amssymb, prl]{revtex4-1}

\usepackage{bm,graphicx, wrapfig}
\usepackage{amsmath,amssymb,latexsym}

\newcommand\mdot{m_\bullet}

\begin{document}

\title{Frame-dragging induced asymmetry in the bending of light near a Kerr black hole} 

\author{S. V.  Iyer}
\email{iyer@geneseo.edu}
\affiliation{Department of Physics \& Astronomy, State University of New York at Geneseo,\\
1 College Circle,
Geneseo, NY 14454.}

\begin{abstract}
The deflection of light's trajectory has been studied in many different spacetime geometries in weak and strong gravity, including the special cases of spherically symmetric static and spinning black holes. It is also well known that the rotation of massive objects results in the dragging of inertial frames in the spacetime geometry.  We present here a discussion of the asymmetry that appears explicitly in the exact analytical expression for the bending angle of light on the equatorial plane of the spinning, or Kerr, black hole.  
\end{abstract}


\maketitle

\section{Introduction}
\label{sec:intro}

The study of gravity's effect on light in general relativity, or gravitational lensing, began with Einstein's prediction in 1913, and the subsequent confirmation by Eddington in 1919, of the bending angle of light in the rather weak gravitational field near the sun. A program to calculate the bending angle of light near a Schwarzschild black hole was started in the late 1950's by Darwin \cite{darwin}, then picked up again for Kerr geometry by Boyer and Lindquist \cite{boy-lind} and many others including Chandrasekhar \cite{chandra}. This rather straightforward analysis begins with the equations of motion for photon orbits, followed by calculations that involve cubic polynomials, and finally resulting in an analytical expression for the bending angle in terms of complete and incomplete elliptic integrals. Calculations for the Schwarzschild case showed early on that as we approach the depths of the gravitational potential, the bending angle exceeds $2\pi$, indicating that multiple looping of a light ray around the center of attraction is possible (see for example, pages 672-678 in \cite{MTW73} or pages 123-134 in \cite{chandra}). While much of this method has been known for decades, the calculation of the exact bending angle for light rays on the equatorial plane of a Kerr black hole was completed only recently\cite{iyerhansen1}. It is important to note here that extensive {\em numerical} calculations and visualizations for light (and particle) orbits in both these geometries are abundant in literature. In this paper, we present an overview of Darwin's method and the analytical results for both Schwarzschild and equatorial Kerr geometries, followed by a comparison of the results and their implication.

\section{Lensing Configuration}
\begin{figure}[htbp] 
\begin{center} 
\includegraphics[width=3in]{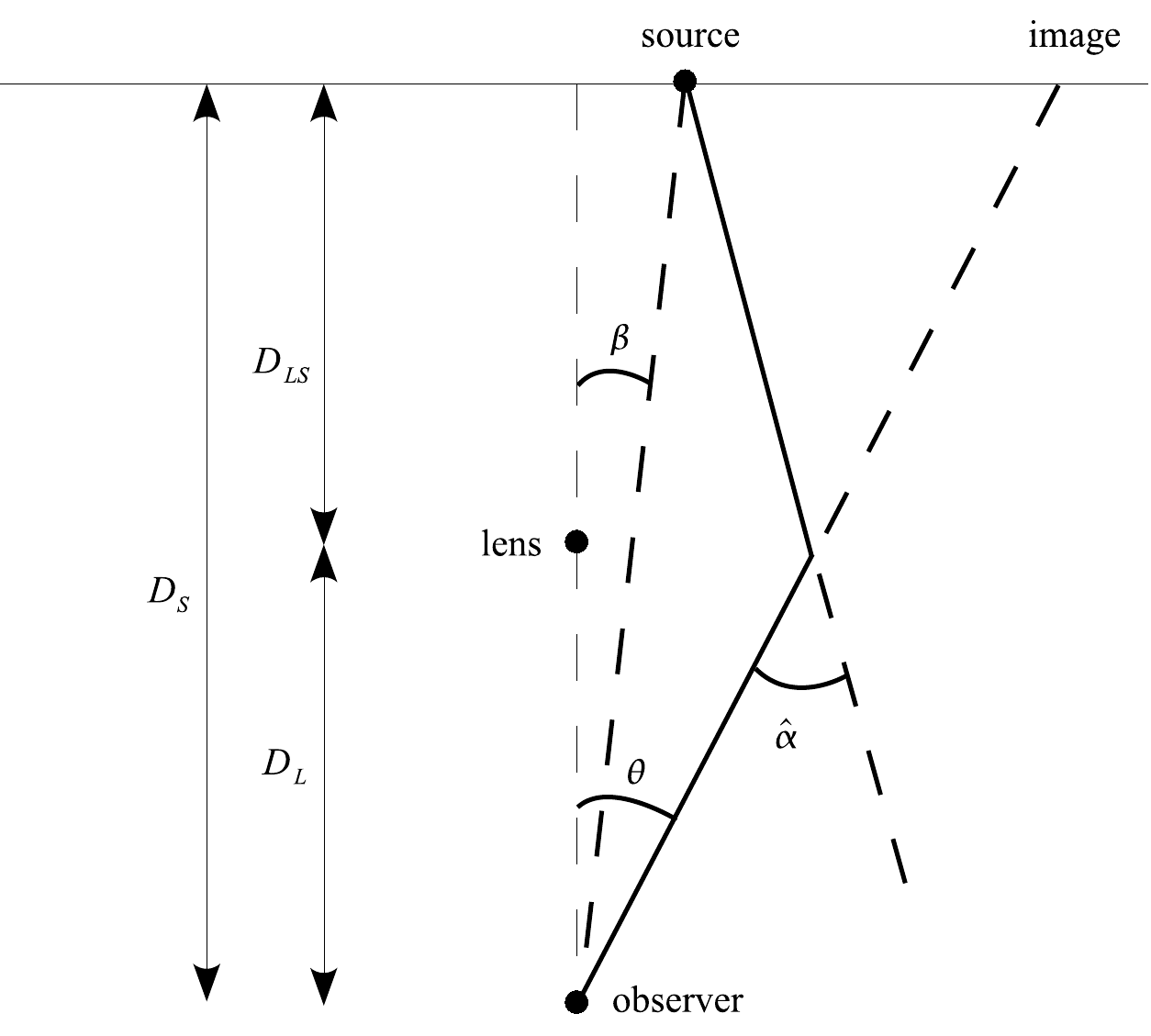}
\caption{\small \sl Thin lens geometry: Source and observer are in the asymptotically flat region, far from the black hole, and on the equatorial plane.} 
\label{LensGeometry} 
\end{center} 
\end{figure}

Consider a standard gravitational lensing situation where a point source and the observer lie in the asymptotically flat region. The lensing geometry is shown in Fig. \ref{LensGeometry}, where it is assumed as usual that the thickness of the lens plane is much smaller than the distances between source and lens $D_{LS}$.  In this view from above the equatorial plane of the black hole, the distance (from the observer) to the lens and source are denoted by $D_{L}$ and $D_{S}$ respectively.  The angular position of the source, $\beta$ angular position of the the image, $\theta$, and the bending angle $\hat{\alpha}$ are also shown in the figure. 

We limit ourselves to the case where the source and the observer are aligned such that light from the source as it goes through the gravitational field and reaches the observer is confined to the equatorial plane of the spinning black hole. Fortunately, because of spherical symmetry, we can consider the same source-lens-observer alignment for the Schwarzschild geometry without loss of generality.  The goal of the program is to obtain analytical results for the bending angle of light that can then be used these to solve for image positions ($\theta$) in the sky.

\section{The Exact Bending Angle for Schwarzschild Geometry}

The line element for the static, spherically symmetric, asymptotically flat vacuum solution of the Einstein equation is given in Schwarzschild coordinates $(t,r,\theta,\phi)$ (with $\theta$ held constant at $\pi/2$ for the equatorial plane) by 
$$ ds^2= -\left(1-\frac{2\mdot}{r}\right)dt^2 + \left(1-\frac{2\mdot}{r}\right)^{-1}dr^2 + r^2\,d\phi^2$$
where $t = c \tau$ and $\mdot=G M/c^2$ (gravitational radius) with $\tau$ physical time and $M$ the physical mass of the black hole.  From the Euler-Lagrange equations for null geodesics, we have (see for example, page 673 in \cite{MTW73}): 
$$\left(\frac{1}{r^2}\frac{dr}{d \phi}\right)^2 +\frac{1-2\mdot/r}{r^2}= \frac{1}{b^2},$$
where $b = |L/E|$ is the impact parameter with $L$ and $E$ the respective angular momentum and energy invariants of the light ray. Setting $u = 1/r$, we can rewrite the above as 
\begin{equation}
\label{cubicSchw}
\left(\frac{du}{d\phi}\right)^2 = 2 \mdot u^3 -u^2 + {1\over{b^2}}.
\end{equation}
From here on, following a detailed analysis (\cite{iyerpetters}) of this cubic polynomial and its roots, we can show that the light ray undergoes a deflection $\hat{\alpha}$, given by:
\begin{equation}
\label{ExactSchwBangle} 
\hat{\alpha}=\pi +4\sqrt{\frac{r_0}{Q}} \left[K(k) - F(\psi,k)\right], 
\end{equation}
where the parameters $Q, \psi$, and $k$ are functions of $\mdot$ and $r_0$ (see \cite{iyerpetters} for details), and $K(k)$ and $F(\psi,k)$ are the complete and incomplete  elliptic integrals of the first kind, respectively.  Furthermore, one can easily verify that in the $\mdot=0$ limit, we obtain $Q=r_0$, $k=0$ and $\psi=\pi/4$, which in turn imply that $K(k)=\pi/2$ and $F(\psi,k)=\pi/4$, giving us {\em zero deflection} as expected.

\section{The Exact Bending Angle for Kerr Geometry}

\begin{figure}[htbp] 
\begin{center} 
\includegraphics[width=3in]{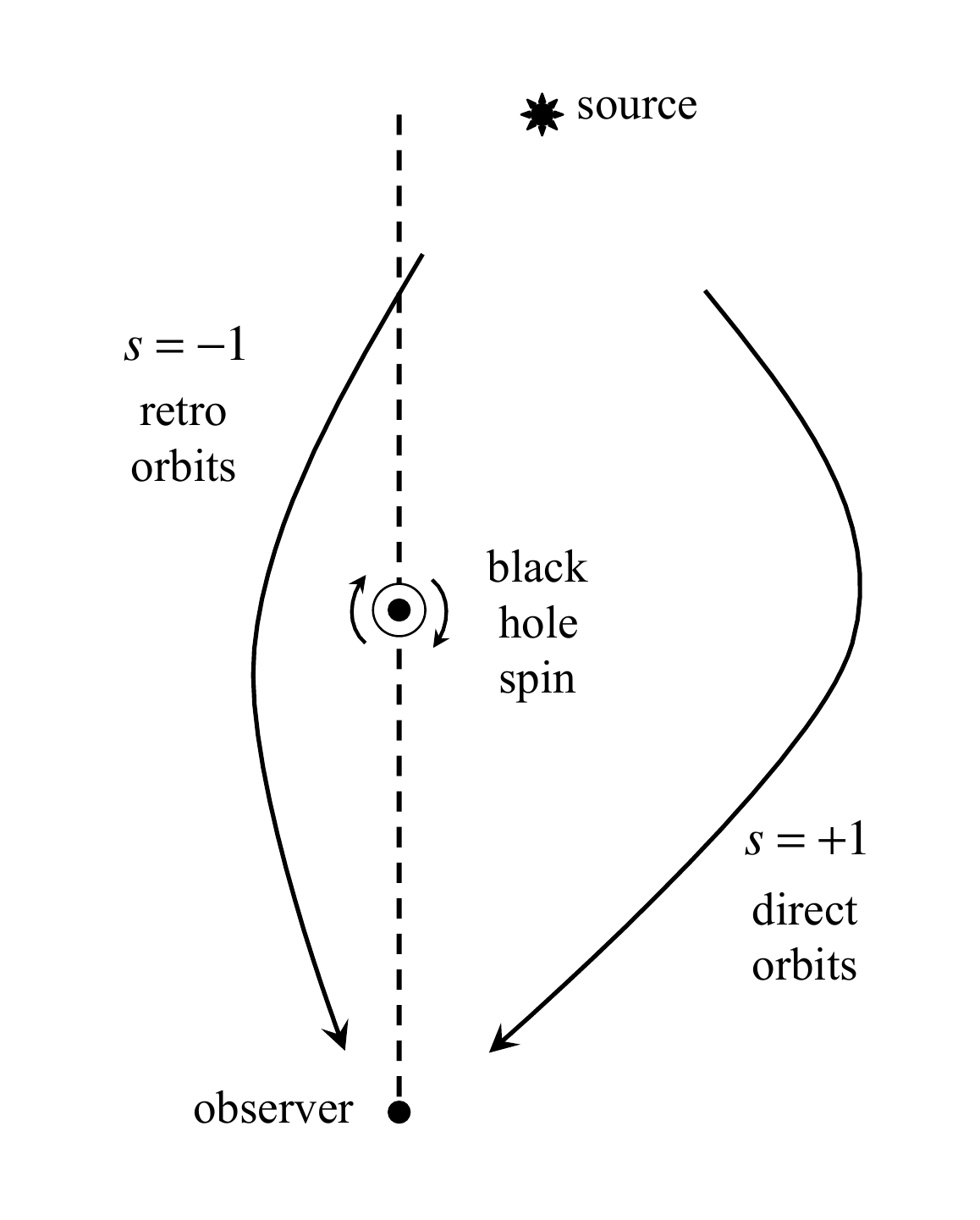}
\caption{\small \sl Sign convention for orbits as viewed from above.  The spin axis points into the page in this figure.} 
\label{OrbitSign} 
\end{center} 
\end{figure}

A similar treatment of photon orbits in Kerr geometry begins with the Kerr line element expressed in the Boyer-Lindquist coordinates (with $\theta=\pi/2$ for the equatorial plane):
$$ds^2 = g_{tt}\,dt^2 + g_{rr}\,dr^2 \ + \ g_{\phi \phi}\,d\phi^2 +  2 g_{t \phi}\, dt d\phi$$
where
\begin{align*}
g_{tt}(r)&=-\left(1-\frac{2\mdot}{r}\right)\\
g_{rr}(r)&=\left(1-\frac{2\mdot}{r}+\frac{a^2}{r^2}\right)^{-1}\\
g_{\phi\phi}(r)&=r^2+a^2+\frac{2\mdot a^2}{r}\\
g_{t\phi}(r)&=\frac{-2\mdot a}{r}
\end{align*}
and where $t = c \tau$ and $\mdot=G M/c^2$, with physical time $\tau$ as before, and the physical mass $M$ of the black hole.  Once again, from the Euler-Lagrange equations for null geodesics, we have
$${\dot{r}}^2=\left[{\cal E}^2+\frac{{\cal E}^2a^2}{r^2}+\frac{2M{\cal E}^2a^2}{r^3}-\frac{4Ma{\cal E}L_z}{r^3}-\frac{L_z^2}{r^2}+\frac{2ML_z^2}{r^3}\right]$$
and
$$\dot{\phi}=\frac{L_z\left(1-\dfrac{2\mdot}{r}+\dfrac{2\mdot a}{r}\dfrac{\cal E}{L_z}\right)}{(r^2-2\mdot r+a^2)},$$
where $\cal{E}$ is the energy invariant and $L_z$ is the component along the black hole spin axis direction of the angular momentum component of the light ray. Couched in the quantity $L_z$ that appears even at this early stage, is the {\em sign} of the light ray's angular momentum component. Note that it is important to treat the direct and retrograde orbits separately as we move forward.  We will use the sign to keep track of whether the light ray is traversing along the direction of frame-dragging or opposite to it by defining
\begin{equation}
\label{ImpactSign}
b_s=s\left|\frac{L_z}{{\cal E}}\right|\equiv s b
\end{equation}
with $s={\rm Sign}(L_z/{\cal E})$ and $b$ is the positive magnitude.  The parameter $s$ is positive for direct orbits and negative for retrograde orbits as shown in Fig. \ref{OrbitSign}.

With $a=J/Mc$, where $J/M$ is the angular momentum per unit mass of the black hole, it is convenient to introduce the following notation:
\begin{equation}
\label{ahatmdot}
\hat{a}=\frac{a}{\mdot}=\frac{Jc}{GM^2}
\end{equation}
limiting ourselves to cases where $0\le \hat{a} \le 1$, with $\hat{a}=0$ being the Schwarzschild limit and $\hat{a}=1$ being extreme Kerr.  Orbits with $s=+1$ and $b_s>0$ will be referred to as direct orbits; and those with $s=-1$ and $b_s< 0$ as retrograde orbits.  

The critical impact parameter that separates inspiralling and emergent photon orbits written as a function of the mass $\mdot$ and spin parameter $\hat{a}$ is given by:
\begin{equation}
b_{sc}=-a+s6\mdot\cos{\left[\frac{1}{3}\cos^{-1}{\left(\frac{-sa}{\mdot}\right)}\right]}
\end{equation}
From a lensing perspective, we are interested in impact parameters ranging from beyond the critical value (SDL) extending all the way to infinity (WDL).  With this in mind, we define a dimensionless quantity
\begin{equation}
b'=1-\frac{s b_{sc}}{b}
\end{equation}
where the quantity $s$ guarantees that the $b'$ stays between $0$ and $1$ for direct and retrograde orbits.  Note that this definition goes over naturally in the Schwarzschild limit
\begin{equation}
1-\frac{s b_{sc}}{b}\quad \xrightarrow{\quad a \rightarrow 0\quad} \quad1-\frac{3\sqrt{3}\mdot}{b}\nonumber,
\end{equation}
where the Schwarzschild critical value $3\sqrt{3}\mdot$ is recovered nicely.

Another important note about the notation used in this paper is in order.  Whenever possible, we have used the same variable to represent corresponding quantities in the Schwarzschild and Kerr cases. For example, we use $\hat{\alpha}$ for the bending angle in both cases. Similarly, the quantity $b$ stands for the magnitude of the impact parameter in both cases, while $b_s$ that carries the orbit sign is introduced only for the spinning case.

Rewriting the equations of motion with this new notation, we have
$${\dot{r}}^2=L_z^2\left[\frac{1}{b^2}+\frac{a^2}{b^2r^2}+\frac{2\mdot a^2}{r^3 b^2}-\frac{4\mdot a}{r^3 b_s}-\frac{1}{r^2}+\frac{2\mdot}{r^3}\right].$$
Next, setting $u=1/r$ as before yields (see \cite{iyerhansen1} for details)
\begin{equation}
\label{cubicKerr}
\left(\frac{du}{d\phi}\right)^2=\frac{\left[1-2\mdot u+a^2 u^2\right]^2}{\left[1-2\mdot u\left(1-a/b_s\right)\right]^2} B(u),
\end{equation}
where the quantity $B(u)$, a cubic polynomial given by
$$B(u)=\left[2\mdot \left(1-\frac{a}{b_s}\right)^2 u^3-\left(1-\frac{a^2}{b^2}\right)u^2+\frac{1}{b^2}\right]$$
Following an analysis similar to the Schwarzschild case, we can show that light's bending angle near a Kerr black hole is given by:
\begin{widetext}
\begin{equation}
\label{ExactKerrBangle}
\hat{\alpha}=-\pi+\frac{4}{1-\omega_s}\sqrt{\frac{r_0}{Q}} \bigg\lbrace\Omega_+ \Bigl[\Pi(n_+,k)-\Pi(n_+,\psi,k)\Bigr]+\Omega_- \Bigl[\Pi(n_-,k)-\Pi(n_-,\psi,k)\Bigr]\bigg\rbrace, 
\end{equation}
\end{widetext}
where the parameters $Q, \psi, k, \Omega_\pm, \omega_0, \omega_\pm$, and $n_\pm$ are all explicit functions of $\mdot, r_0$, and $a$ (see \cite{iyerhansen1} for details), and $\Pi(n_\pm,k)$ and $\Pi(n_\pm,\psi,k)$ are the complete and the incomplete elliptic integrals of the third kind respectively.  

Once again, in the limiting case as $a \rightarrow 0$, it can be shown that $\Omega_+ =1$, $\Omega_-=0$, and $n_+=0$ and the bending angle expression yields 
\begin{eqnarray}
\hat{\alpha}&=&-\pi+4\sqrt{\frac{r_0}{Q}}\left[\Pi(0,k)
-\Pi(\psi,0,k)\right] \nonumber \\
&=& -\pi+4\sqrt{\frac{r_0}{Q}}\left[K(k)-F(\psi,k)\right],
\end{eqnarray}
where $K(k)$ and $F(\psi,k)$ are the complete and incomplete elliptic integrals of the first kind respectively.  Setting $a=0$ {\em first} and then $\mdot=0$ in Eq. \ref{ExactKerrBangle} we recover zero deflection as expected (see \cite{iyerhansen1} for details). A nice feature of this analysis is how these two levels of ``correspondence principles" work in a smooth, and rather elegant, manner; by turning off one at a time, the spin followed by the mass, we are able to see how to step down from the spinning to static to flat geometry.

\section{Weak Deflection Limit}

The weak deflection limit is simply defined as the limit when the impact parameter is large, $b\rightarrow \infty$ (or equivalently, $b'\rightarrow 1$, or the coordinate, $r_0\rightarrow \infty$), i.e., when the light ray is in the asymptotic region, far from the black hole or star. 

For example, the following approximation for the bending angle in the weak deflection limit for the spinning case appeared in 1967 in one of the earliest papers on the Kerr metric by Boyer and Lindquist (see \cite{boy-lind} and references within):
\begin{equation}
\label{boy-lind}
\hat{\alpha}=4\frac{\mdot}{r_0}\left[1+\frac{a}{r_0}\right],
\end{equation}
where the leading term is the well known Einstein bending angle for the static case.  

A straightforward series expansion of our result for the exact bending angle (Eq. \ref{ExactKerrBangle}) in the weak deflection limit yields
\begin{widetext}
\begin{eqnarray}
\label{AlphaKWeak}
\hat{\alpha}=4\left(\frac{\mdot}{b}\right)+\left[\frac{15\pi}{4}
-4s\hat{a}\right] \left(\frac{\mdot}{b}\right)^2
&+&\left[\frac{128}{3}-10\pi s\hat{a}+4\hat{a}^2\right] \left(\frac{\mdot}{b}\right)^3 \nonumber\\
&&\;\;+\left[\frac{3465\pi}{64}-192s\hat{a}+\frac{285\pi}{16}\hat{a}^2-4s\hat{a}^3\right]
\left(\frac{\mdot}{b}\right)^4+\mathcal{O}\left[\left(\frac{\mdot}{b}\right)^5\right]+...
\end{eqnarray}
\end{widetext}
The spin parameter $\hat{a}$ and the sign $s$ appear explicitly in our series expansion. Although it may never be needed, the weak deflection series can be extended to include terms of much higher orders, for different values of $\hat{a}$ and for both direct and retro orbits. One quick check that confirms coefficients term by term is a comparison to the WDL series expansion (see \cite{iyerpetters}) for the Schwarzschild case, by setting $\hat{a}=0$ in the above expansion to obtain
\begin{widetext}
\begin{equation}
\hat{\alpha}(b)=4\left(\frac{\mdot}{b}\right) +  
\frac{15\pi}{4}\left(\frac{\mdot}{b}\right)^2
+\frac{128}{3}\left(\frac{\mdot}{b}\right)^3+\frac{3465\pi}{64}\left(\frac{\mdot}{b}\right)^4+
\mathcal{O}\left[\left(\frac{\mdot}{b}\right)^5\right]+...
\end{equation}
\end{widetext}

And finally, to compare our result with the Boyer-Lindquist approximation, we use our definition of $\hat{a}$ in Eq. \ref{AlphaKWeak} to obtain
\begin{align*}
\hat{\alpha}(b)&=4\left(\frac{\mdot}{b}\right) - 4s\frac{a}{\mdot} \left(\frac{\mdot}{b}\right)^2+ \frac{15\pi}{4}\left(\frac{\mdot}{b}\right)^2+...\\
&=4\frac{\mdot}{b}\left[1 - s\frac{a}{b}\right] +... 
\end{align*}
With $b\rightarrow r_0$ in this limit, and $s$ positive, our result is in agreement with (\ref{boy-lind}); specifically, we, too, do not see a correction to first order in $1/b$. Note that the part involving the quantity $15\pi/4$ appears in the term that goes like the square of ($\mdot/b$), and does not appear in the Boyer-Lindquist approximation.  

\section{Comparisons and Perspective}

\begin{figure}[htp] 
\begin{center} 
\includegraphics[width=3.3in]{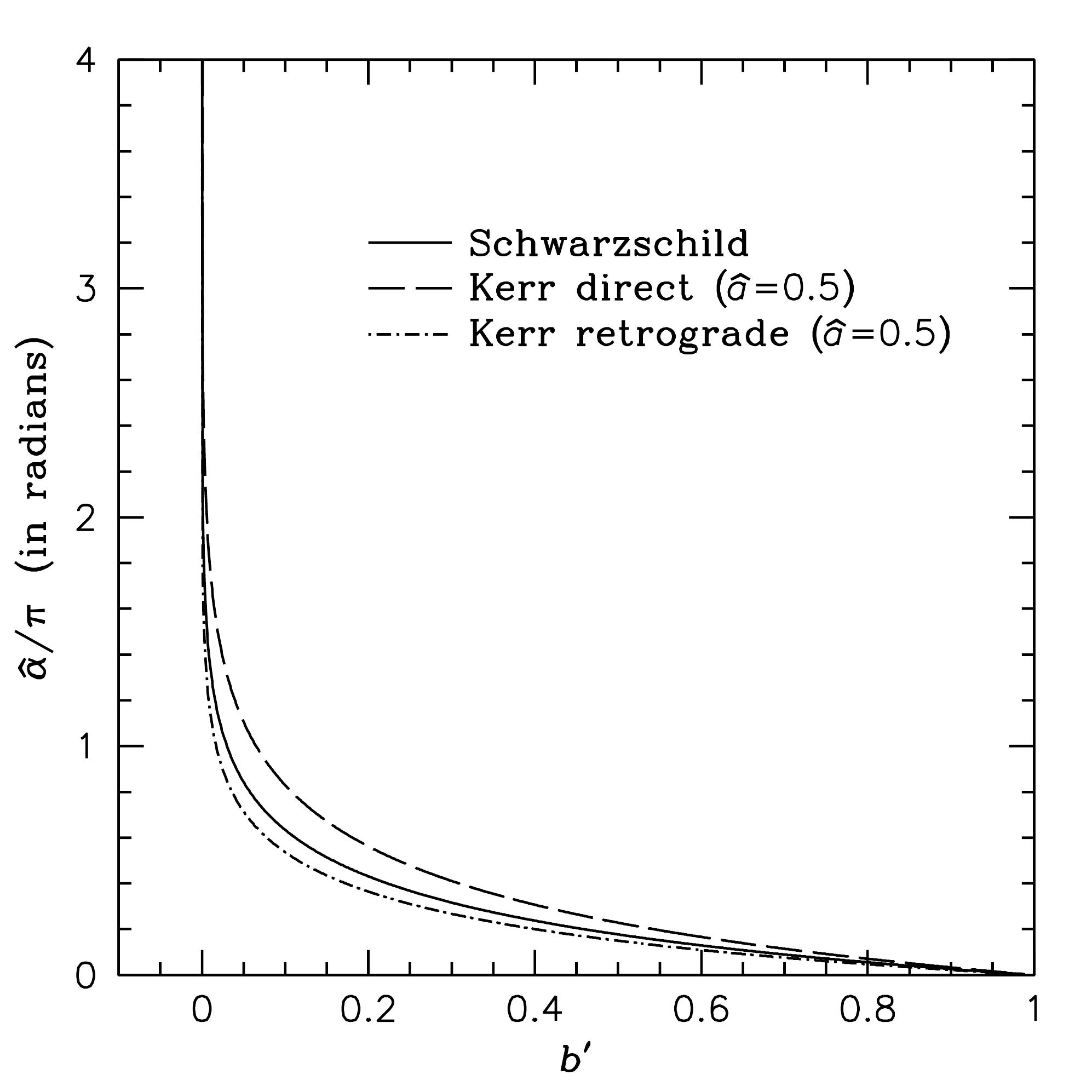}
\caption{\small \sl Bending angles for Schwarzschild and Kerr ($\hat{a}=0.5$) plotted as functions of the normalized impact parameter.  Note that strong deflection limit (SDL) is towards the left of this plot as $b'$ goes to zero, and the weak deflection limit (WDL) is as $b'\rightarrow 1$.} 
\label{ExactKerrp5} 
\end{center} 
\end{figure}

\begin{figure}[htp] 
\begin{center} 
\includegraphics[width=3.3in]{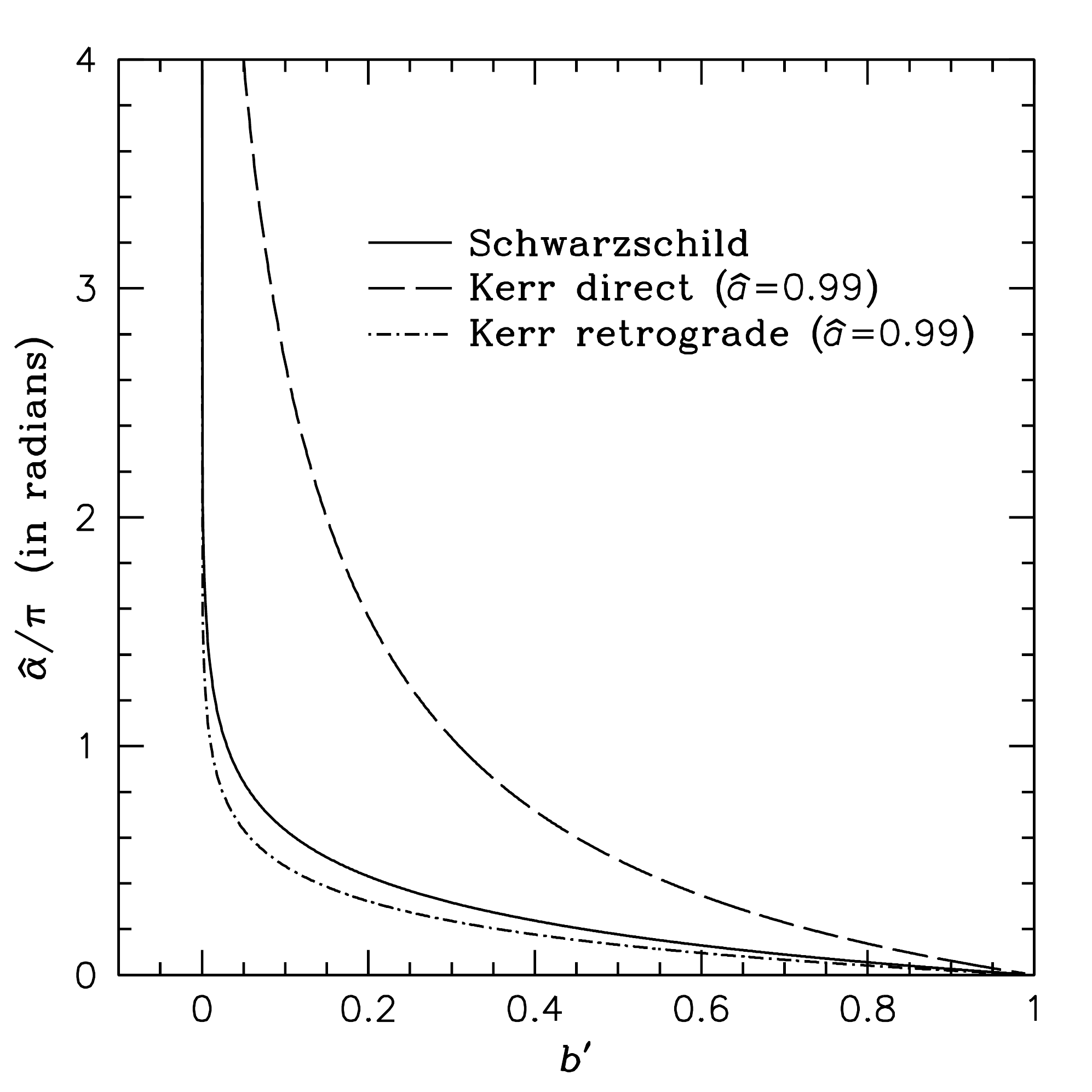}
\caption{\small \sl Bending angles for Schwarzschild and Kerr ($\hat{a}=0.99$) plotted as functions of the normalized impact parameter.  Note that strong deflection limit (SDL) is towards the left of this plot as $b'$ goes to zero, and the weak deflection limit (WDL) is as $b'\rightarrow 1$.} 
\label{ExactKerrp99} 
\end{center} 
\end{figure}

A plot of the bending angle as a function of the impact parameter of the incoming light shows clearly that for orbits just outside a critical impact parameter, the bending angle exceeds $2\pi$ for both the Schwarzschild and Kerr geometries. Images in the sky that correspond to these loop-the-loop paths are referred to as relativistic images. In Figs. \ref{ExactKerrp5} and \ref{ExactKerrp99}, we show Kerr bending angles as dashed curves for two different values, $\hat{a}=0.5$ and $\hat{a}=.99$, and the Schwarzschild bending angle as a solid curve. It is important to note that the introduction of notation in Eq. \ref{ImpactSign} to keep track of the sign of the impact parameter was crucial to seeing the calculation to the end. 

The asymmetry in the amount of deflection on the direct and retrograde sides of the spinning black hole can be seen clearly in the plots: the light ray is being bent more as it's being swirled ``downstream" in the direction of the spin compared to {\em smaller} bending angle as it traverses``upstream." In both direct and retrograde motion, as in the Schwarzschild case, the bending angle exceeds $2\pi$, resulting in multiple loops and the formation of relativistic images.  As a direct consequence of the difference between the direct and the retrograde sides, as seen by an observer in the lensing situation, images in the sky will be offset to different amounts on either side.  This asymmetry is induced by the frame dragging around the spinning black hole; the amount of deflection produced depends on whether the motion of the light ray is in the direction of the black hole spin (direct orbits), or opposite to the spin (retrograde orbits).  Furthermore, we see that for higher spins, the effect is more pronounced resulting in tighter winding of direct orbits with respect to the axis of rotation, and a higher degree of unwinding of retrograde orbits compared to the static case.  As expected, image positions in the sky in gravitational lensing would display asymmetry that can be calculated from the bending angles and corroborated by observation.

While there is no known exact analogy to relativistic frame dragging in Newtonian mechanics, the features described here support our general physical intuition. A natural place to look for analogies is in fluid mechanics. Deflection of a moving boat near a vortex in water or a plane flying near a hurricane are examples of situations where there is rotational dragging; we would expect the impact on the motion along the spin direction to be different from the retrograde side where objects are buffeted by opposing drag. Another somewhat loosely considered analogy in fluid dynamics is the Magnus effect around a spinning baseball or cricket ball that results in a similar asymmetry in airflow around the moving ball. Our final result and the plot of the bending angle shows clearly the asymmetry caused by frame-dragging and the amount of deflection produced on each side: light is deflected more for direct orbits than for retrograde orbits. A rough comparison can also be made to results for time-like geodesics on the equatorial plane near a Kerr black hole (see, for example, pages 326-336 in \cite{chandra}). For circular particle orbits, the magnitude of the angular velocity for direct orbits is lower than that for retrograde orbits. This can be expressed as an asymmetry in the orbital period of the particle, with the period being longer for direct orbits. In basic scattering theory and orbital mechanics, the amount of deflection of particle trajectories depends on the time spent in the force field. Our result may also be thought of as resulting from an asymmetry in the time spent by the photon in the gravitational field, as seen by a distant observer. 

\section{Acknowledgments}

The author acknowledges her colleagues in the Department of Physics and Astronomy at SUNY Geneseo for numerous discussions and their valuable comments.

\end{document}